\documentclass[12pt]{article}
\usepackage[utf8]{inputenc}
\usepackage[english]{babel}
\usepackage[vmargin=0.6in,hmargin={1in,1in},headsep=0.3in,headheight=1in]{geometry}
\usepackage{subfiles}
\usepackage{tikz}
\usepackage{fix-cm}
\usepackage{enumitem}
\usepackage{multicol}
\usepackage{amsmath, amsthm, amssymb, amsfonts,mathrsfs}
\usepackage{paracol}
\usepackage{graphicx}

\usepackage{lineno} 
\allowdisplaybreaks
\newcounter{taggedEquations}
\let\OldTag\tag
\renewcommand*{\tag}[1]{\stepcounter{taggedEquations}\OldTag{#1}}
\usepackage{array}
\usepackage{tabularx}

\usepackage{relsize}
\usepackage{xcolor}
\usepackage{lastpage}
\usepackage{blindtext}
\usepackage[numbers,sort&compress]{natbib}
\usepackage[linktocpage=true]{hyperref}
\hypersetup{
	colorlinks,
	citecolor=blue,
	filecolor=black,
	linkcolor=blue,
	urlcolor=blue
}
\usepackage{url}

\usepackage{graphicx}
\usepackage{caption}
\usepackage{subcaption}
\usepackage{wrapfig}
\usepackage{indentfirst}
\usepackage{hyperref}
\usepackage{soul}
\usepackage{bm}

\usepackage{empheq}
\usepackage{authblk}
\setlist{parsep=0pt,listparindent=\parindent}
\setlist[itemize]{noitemsep, topsep=0pt}
\setlist[enumerate]{noitemsep, topsep=0pt}
\setlist{parsep=0pt,listparindent=\parindent}

\stepcounter{footnote}
\graphicspath{{Images/}{../Images/}}
\usepackage{graphicx}
\graphicspath{ {./images/} }
\usepackage{pgfplots}
\usepackage{extarrows}
\usepackage{mathtools}
\usepackage{tikz}
\usetikzlibrary{decorations.pathmorphing,patterns}
\usetikzlibrary{decorations.pathreplacing,angles,quotes}
\usetikzlibrary{patterns}
\usetikzlibrary{shapes.arrows, fadings}
\usepackage{xfrac} 
\usepackage{relsize}
\usepackage[most]{tcolorbox}
\tikzfading[name=fade down,
  top color=transparent!0, bottom color=transparent!100]
\tikzfading[name=fade up,
  bottom color=transparent!0, top color=transparent!100]

\usepackage{pgfplots}
\pgfplotsset{compat=1.18}
\usetikzlibrary{external}

\usepackage{physics}

\usepackage{enumitem}

\def\beq{\begin{eqnarray}}  
\def\eeq{\end{eqnarray}}

\DeclareFontFamily{U}{matha}{\hyphenchar\font45}
\DeclareFontShape{U}{matha}{m}{n}{
      <5> <6> <7> <8> <9> <10> gen * matha
      <10.95> matha10 <12> <14.4> <17.28> <20.74> <24.88> matha12
      }{}
\DeclareSymbolFont{matha}{U}{matha}{m}{n}

\DeclareMathSymbol{\Lt}{3}{matha}{"CE}
\DeclareMathSymbol{\Gt}{3}{matha}{"CF}

\begin{document}

\title{\Large\textbf{ Where are linearized gauge invariants encoded for plane waves in linearized gravity? }}
\author[1,2]{\large{Ramesh Radhakrishnan }}
\author[3,1]{\large{David McNutt}}
\author[4,1]{\large{Delaram Mirfendereski}}
\author[5,1]{\large{Eric Davis}}
\author[1,2]{\large{William Julius}}
\author[1,2]{\large{Gerald Cleaver}}

\affil[1]{\emph{Early Universe, Cosmology and Strings (EUCOS) Group, Center for Astrophysics, Space Physics and Engineering Research (CASPER)}}
\vspace{1 cm}
\affil[2]{\emph{Department of Physics, Baylor University,  Waco, TX 76798, USA} }
\vspace{1 cm}
\affil[3]{\emph{The Royal Norwegian Naval Academy, Bergen, Norway, }}
\vspace{1 cm}
\affil[4]{\emph{Department of Physics and Astronomy, The University of Texas Rio Grande Valley (UTRGV), USA}}
\vspace{1 cm}
\affil[5]{\emph{Department of Physics, SUNY-Albany, Albany, NY 12222}}
\vspace{0.25cm}

\date{\today}  
\maketitle  
\begin{abstract}

The Newman--Penrose (NP) formalism is traditionally used to analyze the polarization content of gravitational waves, while the gauge-invariant Bardeen formalism provides a complementary, and often simpler, description based on the irreducible scalar, vector, and tensor perturbations of the metric.  
In this work we apply the Bardeen formalism to plane gravitational waves in Minkowski spacetime, computing all scalar, vector, and tensor gauge-invariant variables explicitly and demonstrating that only the two transverse-traceless tensor modes survive, as expected for vacuum waves in general relativity.

We then compare these Bardeen variables with curvature-based invariants constructed using the linearized Cartan--Karlhede (CK) algorithm. Although one might anticipate a correspondence, our analysis shows that the CK invariants do \emph{not} capture the polarization modes: the Weyl tensor possesses only a single non-zero Newman--Penrose scalar and the CK algorithm terminates without producing invariants that distinguish the $\oplus$ and $\otimes$ states. However, by computing invariant quantities obtained from the translational Killing vector fields of the Minkowski background that are retained under linear perturbation, we provide an algorithmic approach that reproduces the same physical tensor degrees of freedom captured by the Bardeen variables. 

\end{abstract}

\par

\pagebreak

\section{Introduction}\label{section:intro}

Modified gravity theories predict modifications to gravitational wave signals, such as changes to the waveform due to the wave–generation mechanism, changes in propagation through altered dispersion relations (arising from modified background geometries), and changes in the number of polarization modes \cite{will2014living,mirshekari2012constraining}.  
The tests of gravity performed using data from Advanced LIGO and Advanced Virgo show that all observed events to date are consistent with General Relativity (GR) \cite{theligoscientificcollaboration2021tests}.  
The planned increase in sensitivity of terrestrial interferometers, new generations of detectors such as the Einstein Telescope and Cosmic Explorer \cite{punturo2010einstein,reitze2019cosmic}, the continued development of pulsar–timing techniques, and future space-based detectors such as LISA \cite{lisa2017l3proposal} will enable increasingly stringent tests of GR.  
When detecting the polarization states of gravitational waves (GW), the presence of any polarization mode beyond the usual ``Plus'' and ``Cross'' tensor modes would imply a violation of GR.  
In fact, any metric theory of gravity in four spacetime dimensions can, in principle, predict up to six polarization modes, as originally classified using the $E(2)$-based Newman--Penrose approach \cite{Eardley-PhysRevD.8.3308}:
\begin{itemize}
    \item 2 scalar modes (1 scalar transverse and 1 scalar longitudinal),
    \item 2 vector modes, and
    \item 2 tensor modes.
\end{itemize}

To check the presence or absence of such modes it is important to evaluate gauge-invariant quantities so that physical observables are not contaminated by gauge artifacts \cite{alves2023testing}.  
The Newman--Penrose (NP) formalism has been used extensively in the literature \cite{Paula_2004,alves2023testing,Eardley-PhysRevD.8.3308} to determine the polarization content of gravitational waves in both General Relativity and modified theories of gravity.  
In this formalism, a tetrad basis $(l^{a}, n^{a}, m^{a}, \bar{m}^{a})$ is used to describe spacetime.  
By projecting the Ricci tensor onto these null tetrad vectors, the ten independent components of the Ricci tensor are encoded in the complex scalars $\Phi_{00}$, $\Phi_{01}$, $\Phi_{02}$ and the real scalars $\Phi_{11}$, $\Phi_{12}$, $\Phi_{22}$, together with $\Lambda$.  
These components are then expressed in terms of NP spin coefficients and curvature scalars using the NP differential equations \cite{Gong_2018}, which determine their evolution.

The Euclidean group $E(2)$ arises naturally in this setting as the little group of the null propagation direction $k^{a}$.  
It consists of all transformations that preserve $k^{a}$ and act as isometries on the two-dimensional screen space orthogonal to $k^{a}$.  
Concretely,
\[
E(2) \simeq SO(2)\ltimes \mathbb{R}^{2},
\]
generated by rotations in the screen space and null rotations about $k^{a}$.  
These transformations encode the gauge freedom of the NP tetrad and determine the transformation properties of the Weyl and Ricci scalars, making $E(2)$ the natural symmetry group underlying the classification of GW polarization modes in GR and modified gravity.  
To identify the polarization modes admitted by a given theory, the authors in \cite{Gong_2018} study how the NP scalars transform under the allowed $E(2)$ operations.  
However, determining polarization content in the NP formalism can be computationally cumbersome.  
An alternate approach is the gauge-invariant Bardeen formalism \cite{bardeen1980gauge,kodama1984cosmological,mukhanov1992theory}, which we utilize in this paper.

To determine and classify the physical effects of these quantities in explicit solutions, one can use invariant functions derived from the underlying geometry.  For example, the event horizon or apparent horizon may not be ideal candidates for defining black hole boundaries, as they are nonphysical, instead physical black hole boundaries can instead be identified using the zero-sets of scalar polynomial invariants (SPI) \cite{Coley_2017}. SPIs are not especially suited to characterizing all properties of a spacetime, and often another set of curvature invariants must be used to completely characterize a spacetime. These invariants are known as Cartan invariants and arise from the Cartan-Karlhede algorithm. This will be discussed in more detail in section \ref{sec:CKalg}, but briefly stated, the Cartan invariants are the components of the curvature tensor and its covariant derivatives in a choice of frame determined by the spacetimes geometry. 

Beyond curvature invariants, there exist several other ways to classify spacetimes. Scalar differential invariants (SDI) also provide a powerful classification tool \cite{Kruglikov_2021}. In contrast, SDI arise from the action of the diffeomorphism group on the space of jets of the metric.  
A \emph{jet} refers to the collection of all partial derivatives of the metric up to a fixed order.  
The $k$--jet $j^{k}g$ encodes the metric components together with all derivatives $\partial^{n} g$ for $0\le n\le k$. These quantities are generally used to study partial differential equations, however, they have been used to classify spacetime geometries, such as Kundt waves \cite{kruglikov2019differential}. Another set of invariant quantities arises from the symmetry properties of spacetimes. For highly symmetric black hole spacetimes, the Killing horizon can be detected using quantities constructed from the Killing vector fields  \cite{debus2024symmetricidealsinvarianthilbert,mcnutt2024detectinghorizonssymmetricblack}. In cases where spacetimes admit Killing vectors with nontrivial orbits, a new set of invariant quantities, called Killing invariants \cite{brown2024killing}, can be used.  These invariants permit a refined sub-classification of spacetimes admitting the same isometry group and are substantially less involved than the full CK algorithm.

From the perspective of explicit classification, the Cartan-Karlhede algorithm is one of the most powerful methods available \cite{Stephani:2003}.  
Using this approach, it is possible to distinguish between black holes and wormholes \cite{mcnutt2021geometric}, and such techniques have even been used to classify simple toy models of gravitational waves in GR \cite{coley2012vacuum}.  
While CK-based approaches apply broadly to exact or approximate solutions, it is more common in linearized gravity to use gauge-invariant observables \cite{khavkine2015local,frob2018approaches}.  
Given that the first approach is always applicable while the latter must be adapted to the specific type of linearized solution, it is of considerable interest to determine what each can and cannot achieve. 

Although the broader six–mode $E(2)$ polarization classification is often discussed in the context of modified theories of gravity, the analysis in this paper is carried out entirely within General Relativity.  Our focus is on the gauge-invariant and geometric structures—Newman-Penrose scalars, Bardeen variables, Cartan-based invariants, and Killing invariants - that characterize gravitational-wave polarization content in GR and provide a foundation for future extensions to modified gravity.

So far, we have used the term, invariant, in an informal manner. It is important here to distinguish between two distinct meanings for the term.  In
one meaning, these are quantities defined from a general metric that are constant along orbits of diffeomorphisms. This definition can be adapted
to the case of general metrics parametrized as linear deviations from a background metric, where they become "linearized gauge invariants". These are the invariants of interest in the theory of gravitational waves, and Bardeen variables, as linearized gauge invariants, are an example. Alternatively for a particular fixed metric, there are all sorts of geometric objects that can be associated to it in a way that does not depend on which coordinates this metric is expressed in. In this second sense, curvature invariants and Killing invariants fall into this second class. Interestingly, within linearized gravity with a Minkowski background, the curvature invariants are also linearized gauge invariants.

\section{Extension of the Bardeen formalism to the flat spacetime limit}

In 1980, Bardeen \cite{PhysRevD.22.1882} proposed the gauge-invariant cosmological perturbation formalism.  
He defined a complete set of gauge-invariant variables for perturbations in a Friedmann–Lemaître–Robertson–Walker (FLRW) background spacetime describing a homogeneous, isotropic, and expanding universe.  
In 2000, Bertschinger \cite{bertschinger2000cosmological} used the $(3+1)$ decomposition technique to classify the degrees of freedom for linearized metric perturbations, and in 2005 Flanagan and Hughes \cite{Flanagan_2005} presented the flat-spacetime limit of Bardeen's formalism.  
This method, called the $(3+1)$ irreducible helicity decomposition, recasts the metric (geometry) or source (matter) tensors into a form that allows one to categorize gravitational waves into radiative, non-radiative, and pure-gauge degrees of freedom.  
For a brief treatment, see Alves \cite{alves2023testing}, who applied this formalism to test gravity by mapping gauge-invariant Bardeen variables to observables such as derivatives of the redshift function.

In this section we present these calculations in full detail to acquaint the reader with the techniques of the formalism, which will later be applied to plane gravitational waves.  The exposition closely follows Flanagan~\cite{Flanagan_2005}, and the notation follows Alves~\cite{alves2023testing}.  We begin by expanding the metric around flat spacetime,
\begin{equation}\label{equation1}
g_{\mu\nu}=\eta_{\mu\nu}+h_{\mu\nu},
\end{equation}
where $\eta_{\mu\nu}$ is the Minkowski metric and $|h_{\mu\nu}|\ll 1$ is a small perturbation.

The metric perturbation $h_{\mu\nu}$ is a symmetric rank--2 tensor,
$h_{\mu\nu}=h_{\nu\mu}$, and may therefore be represented as a $4\times4$ matrix,
\begin{equation}\label{equation2}
h_{\mu\nu}=\begin{bmatrix}
h_{tt} & h_{tx} & h_{ty} & h_{tz}\\
h_{tx} & h_{xx} & h_{xy} & h_{xz}\\
h_{ty} & h_{xy} & h_{yy} & h_{yz}\\
h_{tz} & h_{xz} & h_{yz} & h_{zz}
\end{bmatrix}.
\end{equation}

We decompose $h_{\mu\nu}$ by examining how its components transform under spatial coordinate transformations (e.g.\ rotations).  The component $h_{tt}$ is invariant under spatial rotations and may be written as
\begin{equation}\label{equation3}
h_{tt}=h_{00}\equiv 2\psi,
\end{equation}
showing that $h_{tt}$ transforms as a scalar.

The mixed components $h_{ti}$ behave as a spatial vector field and may be written as the sum of a divergence-free vector $\beta_i$ and the gradient of a scalar $\gamma$:
\begin{equation}\label{equation4}
h_{ti}=h_{0i}=\beta_{i}+\partial_{i}\gamma,
\end{equation}
subject to $\partial_i\beta_i=0$.  
Hence $\beta_i$ has two degrees of freedom (DOF).  
In Cartesian coordinates we can freely raise and lower spatial indices.  

The spatial tensor components decompose as
\begin{equation}\label{equation5}
h_{ij}=h_{ij}^{TT}-2\phi\delta_{ij}
+(\partial_{i}\partial_{j}-\tfrac{1}{3}\delta_{ij}\nabla^{2})\lambda
+\tfrac{1}{2}(\partial_{i}\epsilon_{j}+\partial_{j}\epsilon_{i}),
\end{equation}
where $h_{ij}^{TT}$ is transverse-traceless, satisfying three transverse conditions $\partial_i h_{ij}^{TT}=0$ and one traceless condition
$\delta_{ij}h_{ij}^{TT}=0$, leaving 2 DOF.  The field $\phi$ is a scalar (1 DOF), and $\epsilon_i$ is a divergence-free vector ($\partial_i\epsilon_i=0$), giving 2 DOF.  The remaining scalar $\lambda$ generates a trace-free scalar-type tensor via the differential operator in parentheses: the combination
$(\partial_{i}\partial_{j}-\tfrac{1}{3}\delta_{ij}\nabla^{2})$ is itself traceless, in the sense that
$\delta^{ij}(\partial_{i}\partial_{j}-\tfrac{1}{3}\delta_{ij}\nabla^{2})=0$, independently of the function it acts on.

Thus the full perturbation $h_{\mu\nu}$ is characterized by the scalars $[\psi,\gamma,\phi,\lambda]$ (each with 1 DOF), the vectors $[\beta_i,\epsilon_i]$ (each with 2 DOF), and the tensor $h_{ij}^{TT}$ (2 DOF), for a total of 10 DOF.

Next we examine how these fields transform under an infinitesimal gauge transformation generated by the vector
\begin{equation}\label{equation6}
\xi_{\alpha}=(\xi_t,\xi_i).
\end{equation}
We decompose
\begin{equation}\label{equation7}
\xi_{\alpha}=[A,\,B_i+\partial_i C],
\end{equation}
where $A$ and $C$ are scalars and $B_i$ is a divergence-free vector ($\partial_i B_i=0$).  
The gauge transformation
\[
h_{\mu\nu}\rightarrow h_{\mu\nu}-(\partial_\mu\xi_\nu+\partial_\nu\xi_\mu)
\]
induces the following transformations:

Scalar fields:
\[
\psi\rightarrow\psi-\partial_t A,\qquad
\gamma\rightarrow\gamma-A-\partial_t C,
\]
\[
\phi\rightarrow\phi+\tfrac{1}{3}\nabla^2 C,\qquad
\lambda\rightarrow\lambda-2C.
\]

Vector fields:
\[
\beta_i\rightarrow\beta_i-\partial_t B_i,\qquad
\epsilon_i\rightarrow\epsilon_i-2B_i.
\]

The transverse–traceless tensor is gauge-invariant:
\[
h_{ij}^{TT}\rightarrow h_{ij}^{TT}.
\]

We now construct the gauge-invariant combinations known as the Bardeen variables.  
From the scalar fields we obtain
\begin{equation}\label{equation8}
\Phi=-\phi-\tfrac{1}{6}\nabla^{2}\lambda,
\end{equation}
\begin{equation}\label{equation9}
\Psi=-\psi+\dot{\gamma}-\tfrac{1}{2}\ddot{\lambda},
\end{equation}
each with 1 DOF.  
From the vector fields we obtain the divergence-free combination
\begin{equation}\label{equation10}
E_i=\beta_i-\tfrac{1}{2}\dot{\epsilon}_i,
\end{equation}
with 2 DOF, and the TT tensor
\begin{equation}\label{equation11}
h_{ij}^{TT}=h_{ij}^{TT},
\end{equation}
which already has 2 DOF.

These gauge-invariant quantities represent the physical degrees of freedom: 4 non-radiative modes $(\Phi,\Psi,E_i)$ and 2 radiative modes $(h_{ij}^{TT})$, corresponding to the gravitational-wave polarizations $\oplus$ and $\otimes$.

Flanagan and Hughes \cite{Flanagan_2005} show that substituting these into the Einstein equations and decomposing the matter tensor $T_{\mu\nu}$ yields the field equations
\begin{equation}\label{equation12}
\nabla^{2}\Phi=-8\pi\rho,
\end{equation}
\begin{equation}\label{equation13}
\nabla^{2}\Psi=4\pi(\rho+3p-3\dot S),
\end{equation}
\begin{equation}\label{equation14}
\nabla^{2}E_i=-16\pi S_i,
\end{equation}
\begin{equation}\label{equation15}
\square h_{ij}^{TT}=-16\pi\sigma_{ij}.
\end{equation}
The first three are Poisson-like and thus non-radiative; only \eqref{equation15} is a wave equation, identifying $h_{ij}^{TT}$ as the radiative sector.

In the next section we construct a metric for plane gravitational waves, and in the following section we apply the Bardeen formalism developed here to analyze their polarization content.

\section{Derivation of the metric for plane gravitational waves}

The linearized Einstein equations may be written in wave equation form
\cite{Grøn-Hervik}:
\begin{equation}\label{equation16}
\Box h_{\mu\nu} = -\kappa T_{\mu\nu},
\end{equation}
where $\kappa = 16\pi G / c^{4}$, $G$ is Newton's constant, and $c$ is the
speed of light.  In Lorenz gauge, the linearized Einstein equations
reduce to decoupled wave equations for the metric perturbation.  For a
detailed derivation, see Section~2 of Flanagan and Hughes
\cite{Flanagan_2005}.  In vacuum,
\begin{equation}\label{equation17}
T_{\mu\nu}=0,
\end{equation}
and therefore
\begin{equation}\label{equation18}
\Box h_{\mu\nu}=0.
\end{equation}

A general solution may be written as a superposition of plane waves
\cite{Ta-Pei-Cheng}.  Consider the plane-wave ansatz
\begin{equation}\label{equation19}
h_{\mu\nu}(x)=A_{\mu\nu} e^{i(k_{\alpha}x^{\alpha})},
\end{equation}
or equivalently
\begin{equation}\label{equation20}
h_{\mu\nu}(x)=A_{\mu\nu}\cos(k_{\alpha}x^{\alpha}),
\end{equation}
where $A_{\mu\nu}$ is a constant symmetric polarization tensor and
$k^{\alpha}=(\omega,k_x,k_y,k_z)$ is the wave vector.  Substituting
(\ref{equation19}) into (\ref{equation18}) yields
\begin{equation}\label{equation21}
\Box A_{\mu\nu} e^{i(k_{\alpha}x^{\alpha})}=0,
\end{equation}
which implies
\begin{equation}\label{equation22}
k^{2} A_{\mu\nu} e^{i(k_{\alpha}x^{\alpha})}=0,
\end{equation}
and therefore
\begin{equation}\label{equation23}
k^{2}=k_{\alpha}k^{\alpha}=-\omega^{2}+\vec{k}^{\,2}=0.
\end{equation}
Thus the wave vector is null and $\omega = |\vec{k}|$.  With $c=1$, this
gives $\omega/|\vec{k}|=1$, showing that gravitational waves propagate
at the speed of light.

We impose the Lorenz (harmonic) gauge condition
\begin{equation}\label{equation25}
\partial^{\mu}\bar{h}_{\mu\nu}=0,
\end{equation}
where $\bar{h}_{\mu\nu}=h_{\mu\nu}-\frac{1}{2}h\,\eta_{\mu\nu}$ is the
trace-reversed metric perturbation.  This condition is chosen to
simplify the linearized Einstein equations and does not require
$\partial^{\mu}h=0$.  For a plane-wave solution, the Lorenz gauge
imposes the transversality condition
\begin{equation}\label{equation24}
k^{\mu}A_{\mu\nu}=0.
\end{equation}

The Lorenz gauge does not completely fix the gauge freedom.  One
retains the residual transformation
\[
h_{\mu\nu}\rightarrow h_{\mu\nu}
-(\partial_{\mu}\xi_{\nu}+\partial_{\nu}\xi_{\mu}),
\qquad
\Box \xi_{\mu}=0,
\]
which preserves the Lorenz condition (\ref{equation25}).

Using this freedom, one may choose a gauge in which the temporal
components vanish,
\[
A_{\mu 0}=A_{0 \mu}=0,
\]
and impose the traceless condition
\[
A^{\mu}{}_{\mu}=0.
\]
This defines the transverse--traceless (TT) gauge.

In TT gauge the perturbation satisfies
\begin{equation}\label{equation26}
h_{\mu 0}^{TT}=0,
\end{equation}
\begin{equation}\label{equation27}
\partial^{j}h^{TT}_{kj}=0,
\end{equation}
\begin{equation}\label{equation28}
h^{TT}_{ii}=0.
\end{equation}

For a plane wave propagating in the $+z$ direction,
$k^{\alpha}=(\omega,0,0,\omega)$.  Using $A_{0\nu}=0$ and the
transversality condition (\ref{equation24}), we obtain
\[
\omega A_{3\nu}=0 \quad \Rightarrow \quad A_{3\nu}=A_{\nu 3}=0.
\]
Thus the metric perturbation takes the form
\begin{equation}\label{equation30}
h_{\mu\nu}(t,z)=
\begin{bmatrix}
0 & 0 & 0 & 0\\
0 & A_{xx} & A_{xy} & 0\\
0 & A_{xy} & -A_{xx} & 0\\
0 & 0 & 0 & 0
\end{bmatrix}
\cos[k(t-z)],
\end{equation}
consistent with the standard result \cite{2003gieg.book.....H}.

The two independent tensor polarization states are then
\begin{equation}\label{equation31}
A_{\oplus}^{\mu\nu}
=
h_{\oplus}
\begin{bmatrix}
0 & 0 & 0 & 0\\
0 & 1 & 0 & 0\\
0 & 0 & -1 & 0\\
0 & 0 & 0 & 0
\end{bmatrix},
\end{equation}
and
\begin{equation}\label{equation32}
A_{\otimes}^{\mu\nu}
=
h_{\otimes}
\begin{bmatrix}
0 & 0 & 0 & 0\\
0 & 0 & 1 & 0\\
0 & 1 & 0 & 0\\
0 & 0 & 0 & 0
\end{bmatrix},
\end{equation}
where $h_{\oplus}$ and $h_{\otimes}$ denote the amplitudes of the two
polarization modes.

Using the $\oplus$ polarization for illustration, the metric
perturbation takes the form
\begin{equation}\label{equation33}
h_{\mu\nu}(t,z)
=
h_{\oplus}
\begin{bmatrix}
0 & 0 & 0 & 0\\
0 & 1 & 0 & 0\\
0 & 0 & -1 & 0\\
0 & 0 & 0 & 0
\end{bmatrix}
\cos[k(t-z)].
\end{equation}
An analogous expression holds for the $\otimes$ polarization with
amplitude $h_{\otimes}$.

\section{Bardeen variables calculation for plane gravitational waves} \label{sec:Bardeen}

A high-level description of the construction of gauge-invariant variables in flat spacetime can be found in Jaccard, Maggiore, and Mitsou\cite{Bardeen-Maggiore}.  Here we present the full derivation and apply it explicitly to the case of plane gravitational waves.

From Section~2, the gauge-invariant scalar variables are recalled here
for completeness before being evaluated explicitly for the plane-wave
solution:
\begin{equation}\label{equation34}
\Phi=-\phi-\frac{1}{6}\nabla^{2}\lambda,
\end{equation}
\begin{equation}\label{equation35}
\Psi=-\psi+\dot{\gamma}-\frac{1}{2}\ddot{\lambda},
\end{equation}
the divergence-free vector field is
\begin{equation}\label{equation36}
E_i=\beta_i-\frac{1}{2}\dot{\epsilon}_i,
\end{equation}
and the tensor variable is simply
\begin{equation}\label{equation37}
h_{ij}^{TT}=h_{ij}^{TT}.
\end{equation}
The four gauge degrees of freedom have been removed by the transformation generated by $\xi_\alpha$.

Contracting the spatial decomposition (\ref{equation5}) with $\delta^{ij}$
gives
\begin{equation}\label{equation38}
\phi=-\frac{1}{6}h^{i}{}_{i}.
\end{equation}
From (\ref{equation3}) we have
\begin{equation}\label{equation39}
\psi=\frac{h_{00}}{2}.
\end{equation}
Taking the divergence of (\ref{equation4}) and using
$\partial^{i}\beta_i=0$ yields
\[
\partial^{i}h_{0i}=\nabla^{2}\gamma,
\]
and therefore
\begin{equation}\label{equation40}
\gamma=\nabla^{-2}(\partial^{i}h_{0i}),
\end{equation}
where the inverse Laplacian is defined assuming the boundary condition
that $\gamma\to 0$ at spatial infinity \cite{Bardeen-Maggiore}.

Applying $\partial^{i}\partial^{j}$ to (\ref{equation5}), and noting that the TT piece and the transverse vector piece drop out under double divergence, we obtain
\begin{equation}\label{equation41}
\nabla^{2}\lambda
=-\frac{1}{2}h^{i}{}_{i}
+\frac{3}{2}\nabla^{-2}(\partial^{i}\partial^{j}h_{ij}).
\end{equation}
Substituting (\ref{equation41}) into (\ref{equation34}) gives
\begin{equation}\label{equation42}
\Phi=-\frac{1}{4}\nabla^{-2}(\partial^{i}\partial^{j}h_{ij})
+\frac{1}{4}h^{i}{}_{i}.
\end{equation}

For the vector decomposition, write
\[
h_{0i}=\beta_i+\partial_i\gamma.
\]
Using (\ref{equation40}) gives
\begin{equation}\label{equation43}
\beta_i=h_{0i}-\partial_i\nabla^{-2}(\partial^{j}h_{0j}),
\end{equation}
and
\begin{equation}\label{equation44}
\epsilon_i
=2\nabla^{-2}\!\left[\partial^{j}h_{ij}
-\partial_i\nabla^{-2}(\partial^{k}\partial^{l}h_{kl})\right].
\end{equation}

We now apply these expressions to the plane gravitational wave of
Section~3 with $+$ polarization,
\begin{equation}\label{equation45}
h_{\mu\nu}(t,z)=
\begin{bmatrix}
0 & 0 & 0 & 0\\
0 & 1 & 0 & 0\\
0 & 0 & -1 & 0\\
0 & 0 & 0 & 0
\end{bmatrix}
\cos k(t-z).
\end{equation}
From this we obtain $h^{i}{}_{i}=0$, $h_{0i}=0$, $h_{11}=\cos k(t-z)$,
$h_{22}=-\cos k(t-z)$, and all other $h_{ij}=0$.

The TT-gauge conditions $h_{0i}=0$ and $\partial^{j}h_{ij}=0$ imply
\[
\phi=0,\qquad
\psi=0,\qquad
\gamma=0,\qquad
\nabla^{2}\lambda=0,
\]
and therefore $\Phi=0$ from (\ref{equation42}).  From (\ref{equation35})
we have
\[
\Psi=-\frac{1}{2}\ddot{\lambda}.
\]
But (\ref{equation41}) with $h^{i}{}_{i}=0$ and
$\partial^{i}\partial^{j}h_{ij}=0$ implies
\[
\nabla^{2}\lambda=0.
\]
Assuming $\lambda\to 0$ at spatial infinity, we conclude $\lambda=0$,
and therefore $\Psi=0$.

For the vector sector, applying the TT gauge to
(\ref{equation43}) and (\ref{equation44}) gives
\[
\beta_i=0,\qquad
\epsilon_i=0,
\]
and hence $E_i=0$.

Thus for plane gravitational waves in TT gauge, all scalar and vector
Bardeen variables vanish identically.  The only non-zero components of
the metric perturbation are the transverse--traceless tensor components,
\begin{equation}\label{equation47}
h_{11}= h_\oplus = \cos k(t-z),
\end{equation}
\begin{equation}\label{equation48}
h_{22}= -h_\oplus = -\cos k(t-z).
\end{equation}

Repeating the calculation for the $\otimes$ polarization yields
\[
h_{12}= h_\otimes = \cos k(t-z),
\]
as the single non-zero TT component for that mode.

\section{The Newman--Penrose (NP) formalism}

In this section we take a brief detour into the NP formalism, which will be utilized in Section~6. The NP formalism is a well-known approach to studying gravitational waves in GR.  
This approach is based on the tetrad (or spinor) formalism and was introduced in 1962 \cite{Newman:1961qr}.  
An alternate but related construction \cite{Geroch1973} uses pairs of null directions instead of a full null tetrad.  
The NP formalism is also known as the spin-coefficient method.  
It employs complex linear combinations of the Ricci rotation coefficients, which are related to the spinor affine connection, and it can be used to describe the asymptotic behaviour of gravitational waves, among other applications.  
Spinors also provide a powerful language for describing geometric properties of spacetimes \cite{Penrose:1985bww}, transforming under rotations and Lorentz transformations in a specific and well-controlled way.

In this article, since we are primarily interested in computing the NP Weyl and Ricci curvature scalars, we focus on that part of the NP formalism and illustrate the computation following the example given in \cite{PhysRevD.110.124006}.  
The results in Section~6 are based on this procedure.  
The first step is the choice of a null tetrad---a tetrad of null vectors---denoted  
$(l^{\mu}, n^{\mu}, m^{\mu}, \bar m^{\mu})$, 
where $l^{\mu}$ and $n^{\mu}$ are real null vectors, and $m^{\mu}$ and $\bar m^{\mu}$ are complex conjugates.  
This choice, originally made by Newman and Penrose \cite{Newman:1961qr}, is particularly well suited for spacetimes of Petrov type~D, such as Kerr and Schwarzschild.  
The Petrov classification is reviewed in Chapter~4 of \cite{Stephani:2003}.  
Type~D fields typically occur in the exterior region of a rotating or static black hole and are characterized by mass, angular momentum, and two distinct principal null directions corresponding to ingoing and outgoing null congruences.

The tetrad null vectors can be normalized as follows:
\begin{equation}
    l^{\mu} l_{\mu}=n^{\mu} n_{\mu}=m^{\mu} m_{\mu}=\bar m^{\mu} \bar m_{\mu}=0 ,
\end{equation}
\begin{equation}
    l^{\mu} m_{\mu}=l^{\mu} \bar m_{\mu}=n^{\mu} m_{\mu}=n^{\mu} \bar m_{\mu}=0 ,
\end{equation}
\begin{equation}
    l^{\mu} n_{\mu}=-1 , \qquad m^{\mu} \bar m_{\mu}=1 .
\end{equation}

The components of the metric $g_{\alpha\beta}$ in this tetrad basis are
\begin{equation}
    g_{\alpha\beta}
    =e_{\alpha}^{\mu} e_{\beta\mu}
    =
    \begin{bmatrix}
         0 & -1 &  0 &  0\\
        -1 &  0 &  0 &  0\\
         0 &  0 &  0 &  1\\
         0 &  0 &  1 &  0
    \end{bmatrix},
\end{equation}
where 
\begin{equation}
    e_{1}^{\mu}=l^{\mu}, \qquad
    e_{2}^{\mu}=n^{\mu}, \qquad
    e_{3}^{\mu}=m^{\mu}, \qquad
    e_{4}^{\mu}=\bar m^{\mu}.
\end{equation}

We next compute the spin coefficients.  
The covariant derivatives along the tetrad vectors are defined as
\begin{equation}
    D = l^{\mu}\nabla_{\mu}, \qquad
    \Delta = n^{\mu}\nabla_{\mu}, \qquad
    \delta = m^{\mu}\nabla_{\mu}, \qquad
    \bar{\delta} = \bar m^{\mu}\nabla_{\mu}.
\end{equation}
The NP spin coefficients, which encode the optical and geometric properties of the null congruences, are the tetrad components of the covariant derivatives of the tetrad vectors.  
The twelve standard NP coefficients are 
$\kappa, \sigma, \lambda, \nu, \rho, \mu, \tau, \pi, \epsilon, \gamma, \alpha, \beta$  
(see \cite{PhysRevD.110.124006} for explicit definitions).  

We then compute the NP curvature scalars: the Ricci scalars and the five complex Weyl scalars.  
The NP Ricci scalars encode the ten independent components of the Ricci tensor and are obtained by contracting $R_{\mu\nu}$ with the tetrad vectors $(l,n,m,\bar m)$.  
The four real Ricci scalars are
\begin{equation}
    \Phi_{00}=\frac{1}{2}R_{\mu\nu}l^{\mu}l^{\nu}, \qquad
    \Phi_{11}=\frac{1}{4}\!\left(R_{\mu\nu}l^{\mu}n^{\nu}+R_{\mu\nu}m^{\mu}\bar m^{\nu}\right),
\end{equation}
\begin{equation}\label{PHI-22}
    \Phi_{22}=\frac{1}{2}R_{\mu\nu}n^{\mu}n^{\nu}, \qquad
    \Lambda=\frac{1}{24}R,
\end{equation}
where $R$ is the Ricci scalar.
The three complex Ricci scalars are
\begin{equation}
    \Phi_{01}=\frac{1}{2}R_{\mu\nu}l^{\mu}m^{\nu}, \qquad
    \Phi_{02}=\frac{1}{2}R_{\mu\nu}m^{\mu}m^{\nu}, \qquad
    \Phi_{12}=\frac{1}{2}R_{\mu\nu}n^{\mu}m^{\nu}.
\end{equation}

The five complex NP Weyl scalars are obtained by contracting the Weyl tensor $C_{\mu\nu\alpha\beta}$ with the tetrad vectors:
\begin{equation}
  \Psi_0 = l^{\mu} m^{\nu} l^{\alpha} m^{\beta} C_{\mu\nu\alpha\beta}, \qquad
  \Psi_1 = l^{\mu} n^{\nu} l^{\alpha} m^{\beta} C_{\mu\nu\alpha\beta},
\end{equation}
\begin{equation}
  \Psi_2 = l^{\mu} m^{\nu} \bar m^{\alpha} n^{\beta} C_{\mu\nu\alpha\beta}, \qquad
  \Psi_3 = l^{\mu} n^{\nu} \bar m^{\alpha} n^{\beta} C_{\mu\nu\alpha\beta},
\end{equation}
\begin{equation}\label{PSI-4}
  \Psi_4 = n^{\mu} \bar m^{\nu} n^{\alpha} \bar m^{\beta} C_{\mu\nu\alpha\beta}.
\end{equation}
These five scalars collectively encode the ten independent components of the Weyl tensor.  
Although the NP formalism contains additional structure, the curvature scalars introduced above are the essential ingredients required for the computations that follow in Section~6.

In the NP approach the polarization states can be determined from the NP scalars for particular choices of the coframe \cite{Eardley-PhysRevD.8.3308}. For the plane-waves in linearized gravity, it is always possible to choose a NP frame, $\{ \ell, n, m, \bar{m}\}$ where all of the NP scalars vanish except for the NP scalar $\Psi_4$. Then, the $h_\oplus$ and $h_\otimes$ polarizations can be determined, respectively from the real and imaginary parts of $\Psi_4$: 

\begin{equation}
    h_\oplus = Re(\Psi_4), \quad h_\otimes = Im(\Psi_4).
\end{equation}
However, it is also possible to always apply a boost and spin to produce a new NP frame $\{ \ell', n', m', \bar{m}'\}$ for which $\Psi_4 = 1$. In this frame, one cannot determine the polarization state of the plane wave. This is due to NP scalars being dependent of frame choice, and in some sense are not valid invariants on their own, one must specify the frame in an invariant manner which is achieved by the Cartan-Karlhede algorithm.

\section{The Cartan--Karlhede algorithm for the plane-waves} \label{sec:CKalg}

For generic spacetimes, the Cartan-Karlhede (CK) algorithm uses the geometry of a given spacetime to determine a sufficiently invariantly defined frame, up to isotropy. We will denote the frame components of the Riemann tensor and its covariant derivatives up to the $q$th order by $R^q$. The algorithm is then:
\begin{enumerate}
\item Let $q=0$.
\item Compute $R^q$.
\item Fix the frame as much as possible using frame transformations (spins, boost, rotations, null-rotations).
\item Find the invariance group $H^q$ of the frame which leaves $R^q$ invariant. 
\item Find the number of functionally independent components $t^q$ among the set $R^q$.
\item If $t^q \neq t^{q-1}$ or $dim(H^q) \neq dim(H^{q-1})$ then set $q=q+1$ and go to step $2$.
\item Otherwise the set $\{H^p,t^p,R^p \}$, $p=1,...,q$ classifies the solution and the dimension of the isometry group $I$ of the metric follows from the relation $dim(I) = dim(H^q) + N - t^q$. 
\end{enumerate}

\noindent The group $H^p$ is known as the isotropy group. If we wish to compare two metrics $g$ and $g_o$ for equivalence, we start by completing the above classification for each metric. The list of invariants $R^p$ are Cartan invariants for the spacetime.

Within the context of linearized gravity, when the background is Minkowski spacetime, the curvature tensor and its covariant derivatives are linearized gauge invariant. This follows from the Walker-Stewart lemma since the curvature tensor and the covariant derivatives of the curvature tensor all vanish in Minkowski spacetime \cite{stewart1974perturbations}. Thus the Petrov type condition for a particular Weyl tensor is ensured to be linearized gauge invariant.  While the metric, and hence the choice of null frame, is not linearized gauge invariant, the required scalar quantities from the curvature tensor and its covariant derivatives in the CK algorithm will be linearized gauge invariant, since the Np frame formalism is a way to collect algebraically independent components of the curvature tensor.     

The Cartan-Karlhede (CK) algorithm must be modified somewhat from the standard algorithm used for exact solutions, in order to be applicable to linearized gravity. In particular for a given frame $\{e_a\},~a=1,2,3,4$, the corresponding coframe, $\{\theta^a\}$, is chosen to be dual {\it up to first order}. Thus, if one were to consider the full expansion for the frame and coframe, they would no longer be dual, i.e., $e_a \theta^b \neq \delta_a^b$. 
As we are not choosing a geometrically preferred frame for the whole geometry, it is possible that the linearized-(co)frame may admit additional symmetries than the usual frame chosen from the CK algorithm. For example, the tensor quantities using the linearized frame may admit a non-trivial linear isotropy group whereas the traditional CK frame would have a trivial linear isotropy group.

We will consider a monochromatic plane-wave with the mixed-polarization:
\begin{equation}
    \begin{aligned}
        ds^2 = -dt^2 + [1+ h_{\oplus}] dx^2 + 2[h_{\otimes}] dx\, dy 
        + [1-h_{\oplus}] dy^2 + dz^2 .
    \end{aligned}
\end{equation}

\noindent where $h_{\oplus} \ll 1 $ and $h_{\otimes}\ll 1$ take the form:
\begin{equation}
    \begin{aligned}
        h_{\oplus} = A \epsilon \cos(\Omega(t-z)), \quad 
        h_{\otimes} = B \epsilon \cos(\Omega(t-z)).   
    \end{aligned}
\end{equation}

To implement the CK algorithm we must choose a frame. Instead of the frame in \cite{coley2012vacuum} for which the frame and coframe are dual regardless of linearization, we will use a new approach inspired by the frame 
$\{e_a\}$ used in \cite{shoom2024gravitational}:
\begin{equation}
    \begin{aligned}
        e_1 = \partial_t, \quad 
        e_2 = \left[ 1 - \frac{h_{\oplus}}{2} \right] \partial_x 
        - \frac{h_{\otimes}}{2} \partial_y, \quad 
        e_3 = - \frac{h_{\otimes}}{2} \partial_x 
        + \left[1 + \frac{h_{\oplus}}{2}\right] \partial_y, \quad 
        e_4 = \partial_z.
    \end{aligned}
\end{equation}

\noindent Up to linear order its dual coframe $\{\theta^a\}$ is then
\begin{equation}
    \begin{aligned}
        \theta^1 = dt, \quad 
        \theta^2 = \left[ 1 + \frac{h_{\oplus}}{2} \right] dx 
        + \frac{h_{\otimes}}{2} dy, \quad 
        \theta^3 = - \frac{h_{\otimes}}{2} dx 
        + \left[ 1 - \frac{h_{\oplus}}{2} \right] dy, \quad 
        \theta^4 = dz.
    \end{aligned}
\end{equation}

We will work with a Newman--Penrose coframe:
\begin{equation}
    n = \frac{1}{\sqrt{2}}(\theta^1 - \theta^4), \qquad
    \ell = \frac{1}{\sqrt{2}}(\theta^1 + \theta^4), \qquad
    \bar{m} = \frac{1}{\sqrt{2}} (\theta^2 + i \theta^3), \qquad
    m = \frac{1}{\sqrt{2}} (\theta^2 - i \theta^3).
\end{equation}

\noindent so that the metric, up to first order in $h_{\oplus}$ and $h_{\otimes}$, is: 
\begin{equation}
    ds^2 = - \ell n + m \bar{m} + o(\epsilon^2).
\end{equation}

\noindent Then, computing the Weyl and Ricci Newman-Penrose curvature scalars, 
we find that the Ricci tensor vanishes and the Weyl tensor has one non-zero component: 

\begin{equation}
    \Psi_4 = (A-iB) \Omega^2 \epsilon \cos(\Omega(t-z)).
\end{equation}

\noindent We find that at the first iteration of the algorithm we may apply a boost and a spin,
$$ \ell' = C \ell, \qquad n' = C^{-1} n, \qquad m' = e^{i\theta} m, $$ 
\noindent which gives the transformation rule for $\Psi_4$: 
\begin{equation}
    \Psi_4' = C^{-2} e^{-2i \theta} \Psi_4.
\end{equation}
\noindent Choosing $C = \sqrt{|\Psi_4|}$ and $\theta = \frac{1}{2} arg(\Psi_4)$ we find that $\Psi_4' = 1$.  
The zeroth-order linear isotropy group for the linearized solution is the same as the traditional CK frame, namely the subgroup of null rotations about $\ell$ given as  
\begin{equation}
    \ell' = \ell, \quad n' = n + B \bar{m} + \bar{B} m + \frac{|B|^2}{2} \ell, m' = m + \bar{B}\ell. 
\end{equation}
\noindent where $B$ is a complex-valued parameter. 

It is worth noting that in order to normalize $\Psi_4 = 1$ we must divide by a quantity that vanishes whenever $\Omega (t-z) = \frac{\pi}{2} + n \pi $. At these points one has that the Weyl tensor vanishes but that the covariant derivative of the Weyl tensor is non-vanishing. Thus, one may still fix the boost and spin parameters and fix the frame up to null rotations. In what follows we will only present the case for points where $\Psi_4 \neq 0$.

Carrying on to the next iteration of the CK algorithm, the components of the covariant derivative of the Weyl tensor relative to the linearized frame are:
\begin{eqnarray}
C_{abcd;e} m^a \ell^b m^c  \ell^d \ell^e =  
\frac{\Omega \tan(\Omega(t-z)) \epsilon}
{\sqrt{2} \sqrt{A^2+B^2} \sqrt{\cos(\Omega(t-z))}} .
\end{eqnarray}

\noindent Due to the anti-symmetry in the first pair of indices and the second pair of indices of the Weyl tensor, the linear isotropy group at first order is still the subgroup of null rotations about $\ell$. 
As we have recovered one functionally independent component, the CK algorithm must carry on to second order. At second order, the sole non-zero component in the linearized frame is:

\begin{eqnarray}
C_{abcd;ef} m^a \ell^b m^c  \ell^d \ell^e \ell^f =  
\frac{\Omega^2 \epsilon}{2(A^2+B^2) \cos(\Omega(t-z)) }.
\end{eqnarray}

\noindent The linear isotropy group is unchanged due to the antisymmetry in the first and second pairs of indices in the Weyl tensor. Since the metric functions are dependent on $u = t-z$, all of the resulting invariants computed here are dependent on $u$ alone, and so the sole second order invariant is necessarily functionally dependent on the first order invariant. Furthermore, the components of all higher order covariant derivatives of the Weyl tensor will have the same property, so that these components will not yield new functionally independent invariants.  

The stopping condition for the CK algorithm has been achieved and the algorithm concludes at second order. The set of Cartan invariants consists of the algebraically independent components of the first and second covariant derivative of the Weyl tensor. From this analysis, the linearized plane wave solutions share the same problem as their analogues in the full field equations of general relativity \cite{coley2012vacuum}, namely that the polarization modes amount to a choice of spatial frame rotation, 
and do not arise as a Cartan invariant. 

\section{Killing invariants for the plane-waves}

Interestingly, the polarization modes can be detected directly as Killing invariants \cite{brown2024killing}, which are obtained by projecting the metric perturbation onto perturbed background Killing vector fields. While the background spacetime is Minkowski space, and hence admits ten Killing vector fields,  by perturbing the metric of the background spacetime in linearized gravity, we produce a new metric which has a reduced number of Killing vector fields. In the case of the plane waves, there are five Killing vector fields, two of which belong to the isotropy group. 

In standard GR, Killing vector fields and scalars constructed from them are necessarily coordinate invariant quantities for a given solution. By focusing on Killing vector fields which do not have any fixed points, it is possible to use the Killing vector-fields to algorithmically determine an alternative invariantly defined frame. The algorithm for doing so initially relies on the norms of the Killing vector fields to implement a Gram-Schmidt process to produce a frame adapted to the Killing vector fields. The frame can be further specified by using differential invariants arising from the components of the exterior derivatives of the frame basis elements that lie in the span of the Killing vector fields. For the sake of brevity we will omit the details of the algorithm in this paper and instead refer to \cite{brown2024killing}.

As in the CK algorithm, some modifications must be made to account for the fact that the metric is not linearized gauge invariant in linearized gravity. The existence of the Killing vector fields can be determined from the Cartan Karlhede algorithm from the formula for the dimension of the isotropy group. The symmetries realized by the Killing vector fields can be found explicitly from the following equations \cite[Section 8.4]{Stephani:2003}:
\begin{equation}
    \mathcal{L}_X R^\mu_{\nu \gamma \lambda} = 0, \quad \mathcal{L}_X R^\mu_{\nu \gamma \lambda; \zeta_1} = 0, \quad \ldots, \quad \mathcal{L}_X R^\mu_{\nu \gamma \lambda; \zeta_1 \ldots \zeta_p} = 0 
\end{equation}. 
\noindent Alternatively, one can determine all vector fields generating translational symmetry, by solving for all vector fields that annihilate all Cartan invariants \cite[Section 9.4]{Stephani:2003}:.  

In this sense the translational symmetries are linearized gauge invariant as we will always find three translational isometries from the quantities in the CK algorithm, which are linearized gauge invariant: 
\begin{equation}
    \partial_x, \quad \partial_y \text{ and } \partial_t + \partial_z.
\end{equation} From this perspective, we are able to determine the Killing vector fields, in a restricted class of gauge choices, from linearized gauge invariant quantities. The Killing vector fields determined from the Killing equations are not linearized gauge invariant, and so there are gauge choices where the Killing vector fields will have additional terms. Despite this, knowledge of both the vector fields that generate symmetries, and the Killing vector fields can be used to determine linearized gauge invariant quantities.  

To see this, suppose $\phi$ is the diffeomorphism that maps Minkowski spacetime to the general spacetime $(M, g)$ in the gauge where the Killing vector fields are identical to the symmetry generators, so that under a pullback 
\begin{equation}
    h_{\mu \nu } = (\phi^*g)_{\mu \nu } - \eta_{\mu \nu }.
\end{equation}
then, in Cartesian coordinates, it is possible to find three translational isometries as the Killing vector fields $X_i$, $i = 1,2,3$ , in M such that 
\begin{equation}
    \phi_* X_1 = \partial_x, \quad \phi_* X_1 = \partial_y, \quad \phi_* X_3 = \partial_t + \partial_X.
\end{equation}
\noindent Under a linearized gauge transformation, $\xi$, it follows that the metric is now
\begin{equation}
    g = \eta_{\mu \nu} + \hat{h}_{\mu \nu} = \eta_{\mu \nu} + h_{\mu \nu} + \mathcal{L}_\xi \eta,
\end{equation}
\noindent and the Killing vector fields become
\begin{equation}
    \hat{X}_i = X_i + \epsilon \mathcal{L}_\xi X_i, 
\end{equation} 

In practice, $\hat{X}_i$ can be determined by solving the Killing equations for the linearized gauge transformed metric and using the Lie algebra data to isolate the Lie algebra elements. The inner product of two Killing vector fields yields
\begin{equation}
    < \hat{X}_i, \hat{X}_j > = \eta_{\mu \nu} X_i^\mu X_j^\nu + \epsilon ( h_{\mu \nu} X_i^\mu X_j^\nu  + 2 \eta_{\mu \nu} X_i^\mu \mathcal{L}_\xi X_j^nu + \mathcal{L}_\xi \eta_{\mu \nu} X_i^\mu X_j^\nu). \label{eqn:IP_any_gauge}
\end{equation}
\noindent If $\mathcal{L}_\xi X_i = 0$, for $i=1,2,3$, the inner-products of the Killing vector fields will be unchanged. While if $\mathcal{L}_\xi X_i \neq 0$ for some choice of $i$, then it is possible to determine the relevant part of the vector field $\xi$, from the vector quantity:
\begin{equation}
    \hat{X}_i - X_i = \epsilon \mathcal{L}_\xi X_i = \epsilon X_i^\nu\xi_{,\nu}. \label{eqn:GaugeChange} 
\end{equation}
\noindent The integration functions arising from solving for $\xi^\nu$ will not contribute to the inner product, as these are annihilated by the vector fields $X_i$. 

Using this information, one is then able to eliminate first order contributions that arise from $\xi$ in the inner product. Thus the explicit knowledge of $X_i$ and $\hat{X}_i$ allows for the linearized gauge invariant quantities to be isolated regardless of choice of gauge using equations \eqref{eqn:IP_any_gauge} and \eqref{eqn:GaugeChange} in an algorithmic manner.

For the remainder of this section, we will assume that the Lorentz gauge has been chosen using this process, as this corresponds to the gauge where the Killing vector fields are the symmetry generators. We note that $k^\mu = \partial_t + \partial_z$, so that for any gauge which preserves the Lorentz gauge condition, any inner product with a copy of this vector field is zero. Thus, there are only two translational Killing vector fields that are relevant. For the plane waves we do not need to use the full machinery of the Killing invariants algorithm to determine linearized gauge invariant quantities \cite{brown2026killing}. Instead, we can use the inner-products of the translational Killing invariants to produce linearized gauge invariants that identify the polarization states.

For a plane wave propagating in the $+z$ direction,
\begin{equation}
    h_{ij}(t,z)=h_{ij}(t-z)
\end{equation}
for $i,j\in\{x,y\}$. 

Applying the transverse-traceless (TT) gauge,
\begin{equation}
    h_{xx} + h_{yy}=0,
\end{equation}
and $\partial^i(h_{ij}=0)$,
which implies 
\begin{equation}
    h_{xx}=-h_{yy} \equiv h_{\oplus}(t-z),
\end{equation}
and 
\begin{equation}
    h_{xy}\equiv h_{\otimes}(t-z).
\end{equation}

Background Minkowski spacetime admits constant Killing vectors 
\begin{equation}
    \xi(x)=\partial_x, \qquad 
    \xi(y)=\partial_y.
\end{equation}

Now we are ready to write the Killing invariants $K_{ab}(x)$ as projections of the metric perturbation $h_{\mu\nu}$ onto the background Killing vector fields:
\begin{equation}
    K_{ab}(x)=\xi^{\mu}(a)\xi^{\nu}(b) h_{\mu\nu}(x)
\end{equation}
with $a, b \in \{x,y\}$. For the translational Killing vectors 
$\partial_x$, $\partial_y$, these are
\begin{equation}
    K_{xx} = h_{xx}, \qquad
    K_{yy} = h_{yy}, \qquad
    K_{xy} = h_{xy}.
\end{equation}

Using the $TT$ gauge we get
\begin{equation}
    K_{xx}=h_{\oplus}, \qquad 
    K_{yy}=-h_{\oplus}, \qquad 
    K_{xy}=h_{\otimes},
\end{equation}

\noindent so that the linearized plane waves admitting 
${ \bf X} = \frac{\partial}{\partial_x}$ 
and ${\bf Y} = \frac{\partial}{\partial_y}$ 
as Killing vector fields, yield:
\[
|{\bf X}|^2 = - |{\bf Y}|^2 = 2h_{\oplus}, \qquad 
g({\bf X}, {\bf Y}) = h_{\otimes}.
\]

\noindent Thus, the Killing invariants directly reproduce the same physical tensor degrees of freedom identified by the Bardeen variables discussed in section \ref{sec:Bardeen}.

\section{Discussion and Conclusions}

The analysis presented in this work brings together several complementary
perspectives on gravitational waves in general relativity, highlighting the
distinct but interconnected roles of gauge-invariant perturbation variables,
curvature scalars, and geometric classification schemes. Beginning from the
irreducible $(3+1)$ helicity decomposition of the metric perturbation, we
constructed the full set of Bardeen variables in the flat spacetime limit,
thereby isolating the radiative and non-radiative degrees of freedom in a
manner that is explicitly gauge invariant. This provides a clean starting
point for comparing the perturbative description of gravitational waves with
the curvature-based frameworks used in exact solutions.

For plane waves propagating on a Minkowski background we find that only the
two transverse--traceless tensor modes survive, corresponding to the familiar
$\oplus$ and $\otimes$ polarizations. The scalar and vector Bardeen variables
vanish identically in the transverse--traceless gauge, confirming that the
physical radiative sector of linearized gravitational waves in general
relativity is completely captured by the tensor perturbations.

We then applied the Cartan--Karlhede (CK) algorithm in its linearized form to
the same plane-wave geometry. As expected for pp-wave spacetimes, the Weyl
tensor possesses only a single non-zero Newman--Penrose scalar and the CK
algorithm terminates without generating curvature invariants that distinguish
the two tensor polarization states. Our calculation reproduces this
well-known degeneracy explicitly: although the polarization states are
physically meaningful, they do not appear as Cartan invariants of the
spacetime.

In the specific setting of linearized plane waves on a Minkowski background, this naturally raises the question of where the polarization information is encoded geometrically. We showed that for plane waves the polarization
information is captured by Killing invariants constructed from the relevant
translational Killing vectors of the Minkowski background in the Lorentz gauge. In addition, we provide a procedure to determine linearized gauge invariant quantities, such as the polarization information, for a generic gauge choice using symmetries of the curvature tensor and its covariant derivatives in the linearized spacetime using the linearized plane waves as an example. These invariants
arise by projecting the metric perturbation onto the background symmetries that leave the perturbation unchanged,
and provide coordinate-independent quantities that distinguish the tensor
modes. In this sense the Killing invariants reproduce the same physical tensor
degrees of freedom identified by the Bardeen variables. Thus, for plane
gravitational waves, the perturbative gauge-invariant description and the
symmetry-based invariant description encode the same physical content, while
the curvature-based CK invariants remain insensitive to the polarization
choice.

The present analysis is restricted to linearized gravitational waves
propagating on a flat Minkowski background. In this setting the Killing
symmetries of the background spacetime play a central role in defining the
invariants that encode the polarization states. For more general background
geometries the structure of Killing symmetries may differ, and the
relationship between gauge-invariant perturbations and geometric invariants
may change accordingly. The framework developed here therefore provides a
useful diagnostic tool for investigating how gravitational-wave degrees of
freedom are encoded in different invariant structures.

These observations motivate extending the analysis to modified theories of
gravity. In particular, metric $f(R)$ gravity introduces an additional
propagating scalar degree of freedom in the gravitational-wave sector.
In a companion study we will analyze plane-wave solutions in metric
$f(R)$ gravity using the same framework developed here. In that setting
we will examine whether the additional scalar mode is captured by the
gauge-invariant Bardeen scalar sector, whether it manifests in additional
Newman--Penrose curvature scalars (such as $\Phi_{22}$ or $\Psi_2$,
depending on the background and tetrad choice), and whether Killing
invariants constructed from the background symmetries continue to
provide a direct mapping to the physical gauge-invariant observables.
This will allow us to determine whether the relationship identified here
between Bardeen variables and Killing invariants extends beyond general
relativity or is specific to the tensor sector of linearized gravitational
waves in GR.

Overall, this work provides a unified perspective linking perturbative,
curvature-based, and symmetry-based approaches to characterizing
gravitational-wave degrees of freedom, and clarifies the geometric status of
the radiative tensor modes in the simplest setting of plane waves.

\section*{Acknowledgments}
\noindent DDM would like to thank Igor Khavkine for helpful discussions.

\bibliographystyle{unsrt-phys-eucos}
\bibliography{ref}

\begin{thebibliography}{10}
\expandafter\ifx\csname url\endcsname\relax
  \def\url#1{\texttt{#1}}\fi
\expandafter\ifx\csname doi\endcsname\relax
  \def\doi#1{\burlalt{doi:#1}{http://dx.doi.org/#1}}\fi
\expandafter\ifx\csname urlprefix\endcsname\relax\def\urlprefix{URL }\fi
\expandafter\ifx\csname href\endcsname\relax
  \def\href#1#2{#2}\fi
\expandafter\ifx\csname burlalt\endcsname\relax
  \def\burlalt#1#2{\href{#2}{#1}}\fi

\bibitem{will2014living}
C.~M. Will.
\newblock The confrontation between general relativity and experiment.
\newblock {\em Liv. Rev. Rel.}, 17(4), 2014.
\newblock \doi{10.12942/lrr-2014-4}.

\bibitem{mirshekari2012constraining}
S.~Mirshekari, N.~Yunes, and C.~M. Will.
\newblock Constraining generic lorentz violation and the speed of the graviton
  with gravitational waves.
\newblock {\em Phys. Rev. D}, 85:024041, 2012.
\newblock \doi{10.1103/PhysRevD.85.024041}.
\newblock \href{http://arxiv.org/abs/gr-qc/1110.2720}{{arXiv:1110.2720
  [gr-qc]}}.

\bibitem{theligoscientificcollaboration2021tests}
{The LIGO Scientific Collaboration and the Virgo Collaboration and the KAGRA
  Collaboration}.
\newblock Tests of general relativity with gwtc-3, 2021.

\bibitem{punturo2010einstein}
M.~Punturo et~al.
\newblock The einstein telescope: A third-generation gravitational wave
  observatory.
\newblock {\em Class. Quant. Grav.}, 27(19):194002, 2010.

\bibitem{reitze2019cosmic}
D.~Reitze et~al.
\newblock Cosmic explorer: the us contribution to gravitational-wave astronomy
  beyond ligo.
\newblock {\em {preprint}}, 2019.
\newblock \href{http://arxiv.org/abs/gr-qc/1907.04833}{{arXiv:1907.04833
  [gr-qc]}}.

\bibitem{lisa2017l3proposal}
{LISA Study Team}.
\newblock Laser interferometer space antenna: Mission proposal.
\newblock Technical report, European Space Agency, 2017.
\newblock L3 Mission Proposal.

\bibitem{Eardley-PhysRevD.8.3308}
D.~M. Eardley, D.~L. Lee, and A.~P. Lightman.
\newblock Gravitational-wave observations as a tool for testing relativistic
  gravity.
\newblock {\em Phys. Rev. D}, 8:3308--3321, Nov 1973.
\newblock \doi{10.1103/PhysRevD.8.3308}.

\bibitem{alves2023testing}
M.~E.~S. Alves.
\newblock Testing gravity with gauge-invariant polarization states of
  gravitational waves.
\newblock {\em {preprint}}, 2023.
\newblock \urlprefix\url{https://doi.org/10.48550/arXiv.2308.09178}.
\newblock \href{http://arxiv.org/abs/gr-qc/2308.09178}{{arXiv:2308.09178
  [gr-qc]}}.

\bibitem{Paula_2004}
W.~L. de~Paula, O.~D. Miranda, and R.~M. Marinho.
\newblock Polarization states of gravitational waves with a massive graviton.
\newblock {\em Class. Quant. Grav.}, 21(19):4595–4605, September 2004.
\newblock \doi{10.1088/0264-9381/21/19/008}.

\bibitem{Gong_2018}
Y.~Gong and S.~Hou.
\newblock The polarizations of gravitational waves.
\newblock {\em Universe}, 4(8):85, August 2018.
\newblock \doi{10.3390/universe4080085}.

\bibitem{bardeen1980gauge}
J.~M. Bardeen.
\newblock Gauge-invariant cosmological perturbations.
\newblock {\em Phys. Rev. D}, 22:1882--1905, 1980.
\newblock \doi{10.1103/PhysRevD.22.1882}.

\bibitem{kodama1984cosmological}
H.~Kodama and M.~Sasaki.
\newblock Cosmological perturbation theory.
\newblock {\em Prog. Theor. Phys. Supp.}, 78:1--166, 1984.
\newblock \doi{10.1143/PTPS.78.1}.

\bibitem{mukhanov1992theory}
V.~F. Mukhanov, H.~A. Feldman, and R.~H. Brandenberger.
\newblock Theory of cosmological perturbations.
\newblock {\em Phys. Rep.}, 215(5-6):203--333, 1992.
\newblock \doi{10.1016/0370-1573(92)90044-Z}.

\bibitem{Coley_2017}
A.~Coley and D.~McNutt.
\newblock Identification of black hole horizons using scalar curvature
  invariants.
\newblock {\em Class. Quant. Grav.}, 35(2):025013, December 2017.
\newblock \doi{10.1088/1361-6382/aa9804}.

\bibitem{Kruglikov_2021}
B.~Kruglikov and E.~Schneider.
\newblock Differential invariants of kundt spacetimes.
\newblock {\em Class. Quant. Grav.}, 38(19):195017, September 2021.
\newblock \doi{10.1088/1361-6382/abff9c}.

\bibitem{kruglikov2019differential}
B.~Kruglikov, D.~McNutt, and E.~Schneider.
\newblock Differential invariants of kundt waves.
\newblock {\em Class. Quant. Grav.}, 36(15):155011, 2019.

\bibitem{debus2024symmetricidealsinvarianthilbert}
S.~Debus and A.~Kretschmer.
\newblock Symmetric ideals and invariant hilbert schemes.
\newblock {\em Journal of Algebra}, 2025.

\bibitem{mcnutt2024detectinghorizonssymmetricblack}
D.~McNutt and E.~Schneider.
\newblock Detecting horizons of symmetric black holes using relative
  differential invariants.
\newblock {\em Class. Quant. Grav.}, 42(10):105011, 2025.

\bibitem{brown2024killing}
C.~Brown, M.~Gorban, W.~Julius, R.~Radhakrishnan, G.~Cleaver, and D.~McNutt.
\newblock Killing invariants: An approach to the sub-classification of
  geometries with symmetry.
\newblock {\em Gen. Rel. Grav.}, 56(8):92, 2024.

\bibitem{Stephani:2003}
H.~Stephani, D.~Kramer, M.A.H. MacCallum, C.~Hoenselaers, and E.~Herlt.
\newblock {\em Exact Solutions of Einstein's Field Equations}.
\newblock Cambridge University Press, New York, NY, USA, 2003.

\bibitem{mcnutt2021geometric}
D.~D. McNutt, W.~Julius, M.~Gorban, B.~Mattingly, P.~Brown, and G.~Cleaver.
\newblock Geometric surfaces: An invariant characterization of spherically
  symmetric black hole horizons and wormhole throats.
\newblock {\em Phys. Rev. D}, 103(12):124024, 2021.

\bibitem{coley2012vacuum}
A.~Coley, D.~McNutt, and R.~Milson.
\newblock Vacuum plane waves: Cartan invariants and physical interpretation.
\newblock {\em Class. Quant. Grav}, 29(23):235023, 2012.

\bibitem{khavkine2015local}
I.~Khavkine.
\newblock Local and gauge invariant observables in gravity.
\newblock {\em Class. Quant. Grav}, 32(18):185019, 2015.

\bibitem{frob2018approaches}
M.~B. Fr{\"o}b, T.~P. Hack, and I.~Khavkine.
\newblock Approaches to linear local gauge-invariant observables in
  inflationary cosmologies.
\newblock {\em Class. Quant. Grav}, 35(11):115002, 2018.

\bibitem{PhysRevD.22.1882}
J.~M. Bardeen.
\newblock Gauge-invariant cosmological perturbations.
\newblock {\em Phys. Rev. D}, 22:1882--1905, Oct 1980.
\newblock \doi{10.1103/PhysRevD.22.1882}.

\bibitem{bertschinger2000cosmological}
E.~Bertschinger.
\newblock Cosmological perturbation theory and structure formation.
\newblock {\em {preprint}}, 2000.
\newblock \href{http://arxiv.org/abs/astro-ph/0101009}{{arXiv:0101009
  [astro-ph]}}.

\bibitem{Flanagan_2005}
É.~É. Flanagan and S.~A. Hughes.
\newblock The basics of gravitational wave theory.
\newblock {\em New J. Phys.}, 7:204–204, September 2005.
\newblock \doi{10.1088/1367-2630/7/1/204}.

\bibitem{Grøn-Hervik}
Ø. Grøn and S.~Hervik.
\newblock {\em Introduction to Einstein’s Theory of Relativity}.
\newblock Springer New York, NY, 2020.

\bibitem{Ta-Pei-Cheng}
T.~P. Cheng.
\newblock {\em Relativity, Gravitation and Cosmology:A Basic Introduction, 2nd
  Edition}.
\newblock Oxford University Press, Inc., NY, 2010.

\bibitem{2003gieg.book.....H}
J.~B. {Hartle}.
\newblock {\em {Gravity : an introduction to Einstein's general relativity}}.
\newblock Cambridge University Press, 2003.

\bibitem{Bardeen-Maggiore}
M.~Jaccard, M.~Maggiore, and E.~Mitsou.
\newblock Bardeen variables and hidden gauge symmetries in linearized massive
  gravity.
\newblock {\em Phys. Rev. D}, 87:044017, Feb 2013.
\newblock \doi{10.1103/PhysRevD.87.044017}.

\bibitem{Newman:1961qr}
E.~Newman and R.~Penrose.
\newblock {An Approach to gravitational radiation by a method of spin
  coefficients}.
\newblock {\em J. Math. Phys.}, 3:566--578, 1962.
\newblock \doi{10.1063/1.1724257}.

\bibitem{Geroch1973}
R.~Geroch, A.~Held, and R.~Penrose.
\newblock A space--time calculus based on pairs of null directions.
\newblock {\em J. Math. Phys.}, 14(7):874--881, 1973.
\newblock \doi{10.1063/1.1666410}.

\bibitem{Penrose:1985bww}
R.~Penrose and W.~Rindler.
\newblock {\em {Spinors and Space-Time}}.
\newblock Cambridge Monographs on Mathematical Physics. Cambridge Univ. Press,
  Cambridge, UK, 4 2011.

\bibitem{PhysRevD.110.124006}
P.~Jizba and K.~Mudru\ifmmode~\check{n}\else \v{n}\fi{}ka.
\newblock Newman-penrose formalism and exact vacuum solutions to conformal weyl
  gravity.
\newblock {\em Phys. Rev. D}, 110:124006, Dec 2024.
\newblock \doi{10.1103/PhysRevD.110.124006}.

\bibitem{stewart1974perturbations}
J.~M. Stewart and M.~Walker.
\newblock Perturbations of space-times in general relativity.
\newblock {\em Proc R Soc Lond A Math Phys Sci}, 341(1624):49--74, 1974.

\bibitem{shoom2024gravitational}
A.~A. Shoom.
\newblock Gravitational faraday and spin-hall effects of light: Local
  description.
\newblock {\em Phys Rev. D}, 110(2):024029, 2024.

\bibitem{brown2026killing}
C.~Brown, G.~Cleaver, and D.~McNutt.
\newblock Killing invariants in the weak-field regime.
\newblock {\em in preparation}, 2026.

\end{thebibliography}

\end{document}